\documentclass[aps,prl,twocolumn,showpacs,preprintnumbers,amsmath,amssymb]{revtex4}
\usepackage[]{graphicx}
\newcommand{\be}{\begin{equation}}
\newcommand{\ee}{\end{equation}}
\newcommand{\bea}{\vspace{0.25cm}\begin{eqnarray}}
\newcommand{\eea}{\end{eqnarray}}

\def\PRL{{Phys. Rev. Lett.} }
\def\PRA{{Phys. Rev.} A }

\newcommand{\ket}[1]{\mbox{\ensuremath{|#1\rangle}}}

\begin{document}

\title{Experimental realization of Counterfactual Quantum Cryptography }

\author{G. Brida, A. Cavanna, I. P. Degiovanni, M. Genovese, P. Traina}
\affiliation{INRIM, Strada delle Cacce 91, Torino 10135, Italy}

\begin{abstract}
In counterfactual QKD information is  transfered, in a secure way, between Alice and Bob even when
no particle carrying the information is in fact transmitted between them. In this letter we fully
implement the scheme for counterfactual QKD proposed in [T. Noh, \PRL \textbf{103}, 230501
(2009)], demonstrating for the first time that information can be transmitted between two parties
without the transmission of a carrier.
\end{abstract}
 \maketitle

Quantum Key Distribution (QKD) is a method for transmitting a secret key between two partners
(usually named Alice and Bob) by exploiting quantum properties of light. The most important
characteristic of this idea is that the secrecy of the generated key is guaranteed by the very laws
of nature, i.e. by the properties of quantum states \cite{gis,scar}. In the last decade QKD is moving
from laboratories to become a mature technology for commercialization \cite{l}; communications over
more than 100 km having been achieved both in fiber \cite{fib}
 and open air \cite{op}.

However, beyond its commercial interest QKD represents also a fruitful test bed of concepts and
ideas blossoming from quantum information theory and studies on foundations of quantum mechanics
\cite{gis,scar,mg,ekert,v,bostroem,lucamarini}.

In this sense a very interesting scheme was recently presented on this journal by Noh \cite{no},
who suggested a QKD protocol (usually called N09) where the information is transmitted, in a secure
way, between Alice and Bob even when no particle carrying the information is in fact transmitted
between them. In essence the scheme exploits a counterfactual measurement, for this reason it is
also known as Counterfactual QKD (CQKD).

The counterfactual measurement, which relies on fundamental
properties of quantum mechanics, is a typical example of
interaction-free measurement that detects the state of an object
without an interaction occurring between it and the measuring
device. One of the most widely known application of counterfactual
measurement can be found in the Elitzur–-Vaidman bomb-testing problem
\cite{bomb}, a well known thought experiment successfully
experimentally implemented in the nineties \cite{bombexp}.

CQKD \cite{no} challenges the usual paradigma requiring an effective
transmission of a signal carrier (usually a photon) between the two
parties exchanging information and therefore represents a very
important conceptual development paving the way to further studies.

In Ref. \cite{sun} a more efficient and complicated CQKD was
proposed, whereas security issues of the N09 protocol were
considered in Ref. \cite{yin}, where it was proved its unconditional
security by considering its equivalence to an entanglement
distillation protocol. Finally, very recently, a security proof for
intercept-resend attacks in realistic situation (non unit detector
efficiency and presence of dark counts) was provided \cite{zhang}.

A first attempt to realize experimentally Noh's scheme is reported
in Ref. \cite{las}. However, this set up missed the key element of
CQKD, since the photon was indeed transmitted between Alice and Bob.

In this letter we fully implement the N09 CQKD scheme, demonstrating for the first time that
information can be transmitted between two parties without the transmission of an information carrier.



To explain the principle of the proposed protocol, we describe an alternative version, absolutely equivalent to
the original one \cite{no},  which is shown in Fig.1 (a).

Alice randomly rotates the single photon
polarization (which originally is to be assumed horizontal) by means of a half wave plate (HWPA), either by $0$ (bit value
"0") or by $\pi/2$ (bit value "1").
Then, the photon enters one port of a $50:50$ beam splitter (BS), which is the first
element of a Michelson interferometer. After BS, according to
the polarization, the photon is in one of the two orthogonal states:
\bea | \phi_0 \rangle &= (| 0 \rangle_A | H \rangle_B + i |H \rangle_A | 0 \rangle_B ) / \sqrt{2}
\\ |\phi_1 \rangle &= (| 0 \rangle_A | V \rangle_B + i |V \rangle_A | 0 \rangle_B ) / \sqrt{2} \eea
 The path A of the interferometer (containing an optical delay OD and a mirror) is inside Alice's sector, while path B
 reaches Bob's one.




Bob randomly selects one of the two polarisations and detects the photon in this polarisation allowing the photon
in the complementary polarisation to fly back to Alice's site. This is achieved exploiting the HWPB and the Polarizing beam splitter (PBS). In particular, as the PBS addresses the $\ket{V}$ photon towards D2, while \ket{H} photon is sent towards the mirror (M), rotations of the polarization of $0$ and $\pi/2$ induced by the HWPB correspond to the detection of $\ket{V}$ and $\ket{H}$ photon state by D2. If the photon is not detected by D2 but reflected back by M
it passes through the HWPB in the selected position, thus the photon gains back its original polarization state  interfering with itself at BS at Alice's site and, for a proper tuning of
the optical delay OD, it deterministically exits in D0.

When Alice and Bob select complementary polarization rotations, then either the photon is transmitted by
BS and
detected by Bob at D2 with 50\% probability (since its polarization at PBS is vertical), or it is reflected in path A and consequently detected by D0 or D1 with equal probability (25\%).

After the detection is completed Alice and Bob can communicate each other whether or not each of
the detectors clicked. If clicked either D0 or D2, with the purpose of detecting the intervention
of an eventual eavesdropper, they announce both the detected and the initial polarization state. If
D1 clicks Alice compares the initial and final polarization states: if they are consistent she does
not reveal any information, otherwise she announces her result. Alice and Bob can then establish a common key by using only the events when the photon was detected
at D1 with the correct polarization.

The only apparent difference between the scheme discussed here and the original proposal in Ref \cite{no} is in the apparatus used by Bob to detect the photon at D2. Nonetheless the one shown accomplishes exactly the same task, thus
the two schemes should be considered absolutely equivalent.

 The very interesting point of this scheme is
 that the selection of events only at detector D1 correspond to
 photons that have traveled path A, i.e. never exited Alice's sector.
Therefore, the task of creating a secret key has been accomplished
without any photon carrying the information having been  outside
Alice's laboratory.

In the following we present the results of our equivalent implementation
of the protocol which is completely analogous to the one of Fig. 1(a), but it is based on a Mach-Zehnder
interferometer instead of
a Michelson interferometer.

\begin{figure}
   \begin{center}
   \begin{tabular}{cc}
   \includegraphics[scale=0.43]{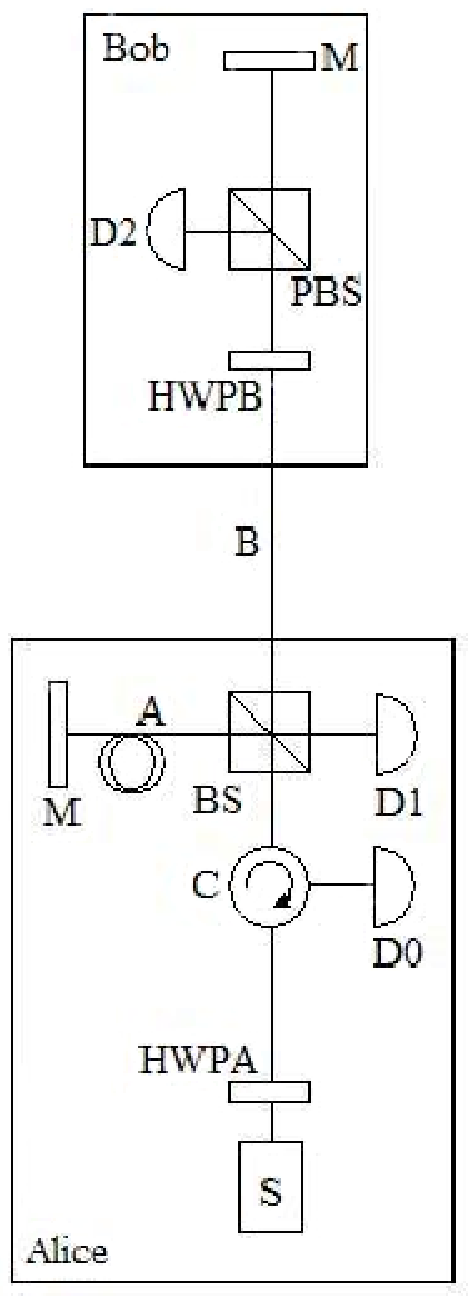}&
   \includegraphics[scale=0.2]{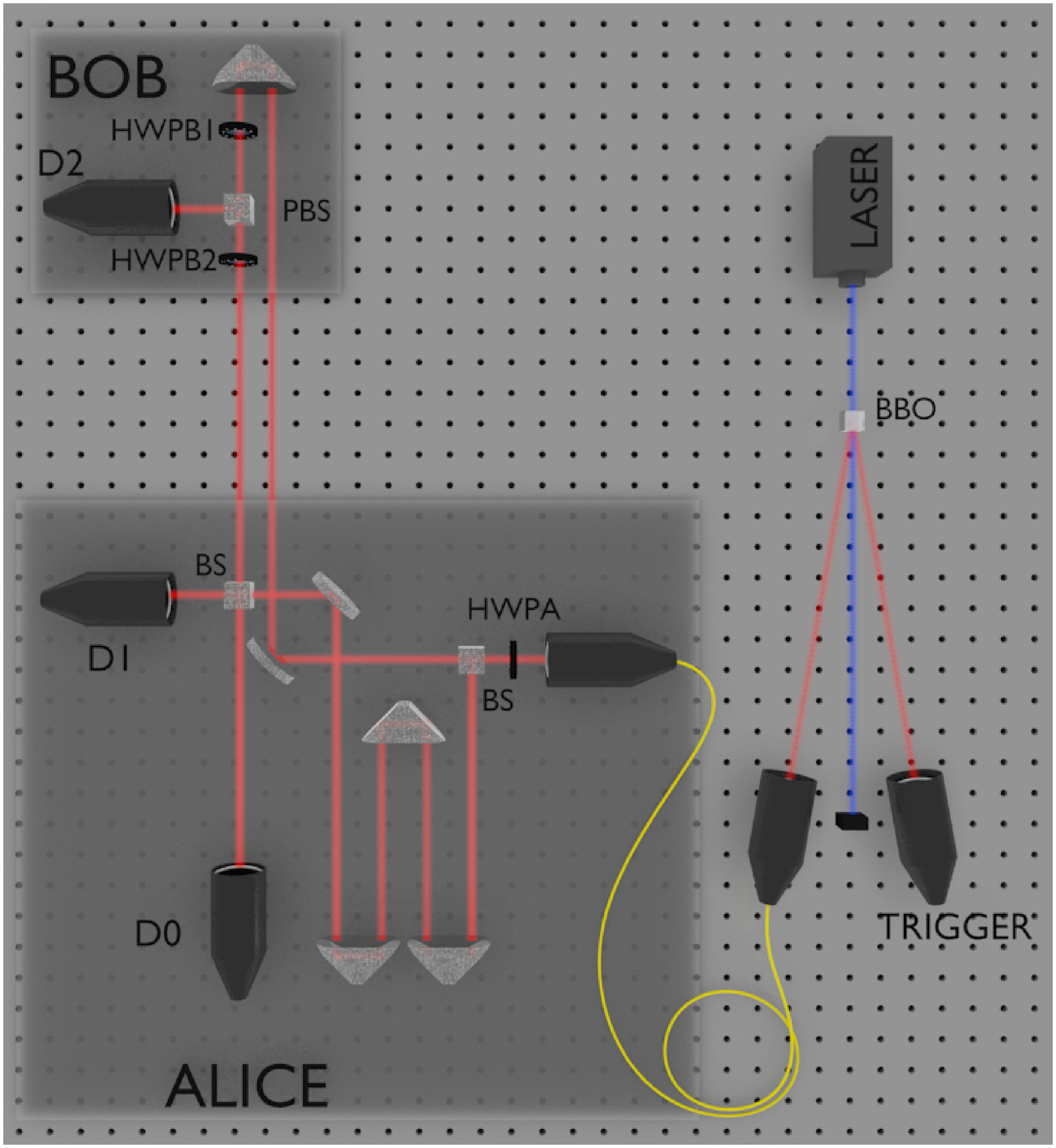}\\
   (a)& (b)
   \end{tabular}
   \end{center}
   \caption[example2]
   { \label{fig:setup}
   (a): scheme of the setup for Counterfactual QKD experiment equivalent to the one proposed by by Noh \cite{no}. (b) Setup of the implemented version of the protocol.
   }
   \end{figure}

 In our experimental set up, as shown in Fig. 1(b), a heralded single photon source
exploiting parametric down-conversion (PDC) is used
: a
100 mW laser emitting at 406 nm in continuous-wave regime pumps a type-I BBO 
crystal producing
degenerate PDC at 812 nm. The emission of the PDC photons is slightly non-collinear
 corresponding to an emission angle of approximately $3^\circ$ with respect to the pump direction.
 The heralding photon after passing through
a 10 nm bandwidth interferential filter and a 4 mm wide pinhole is
coupled to a multi-mode fiber and addressed to the trigger detector.
The heralded photon, to be used as our true single photon state, is
selected by an interferential filter (1 nm FWHM
) and coupled to a single mode fiber leading to the
input of the interferometer.

The latter is a balanced Mach-Zehnder Interferometer (MZI) in which
each arm has an adjustable trombone prism. One of the two arms is
entirely included in Alice's site, while the other contains both the
quantum channel and Bob's site, the latter being composed by a PBS
between two half-wave plates (HWPB1, HWPB2) and D2 detector.

The balance of the interferometer is guaranteed by a
closed--loop piezo--electric movement system, which stabilizes the
position of one of the trombones regulating the length difference between the two optical paths
inside the MZI with nanometric resolution.

The outputs of the interferometer, after spatial selection via $1$ $mm$ diameter-wide irises, are then coupled in multi-mode
fiber with no further spectral selection and all the
signals (including the heralding photons and D2 clicks) are
revealed by Single Photon Avalanche Detectors (SPADs) with a
$\approx 60 \%$ detection efficiency at 812 nm.

Coincidence and time-tag analysis of the incoming signals are performed by means of PicoQuant
HydraHarp 400 multichannel picosecond event timer. All the reported data were acquired in measurements of 20 seconds.
Our results show good agreement with the theoretical predictions and represent a proof of principle of
the experimental feasibility of CQKD.

\begin{figure}
   \begin{center}
   \begin{tabular}{c}
   \includegraphics[scale=0.45]{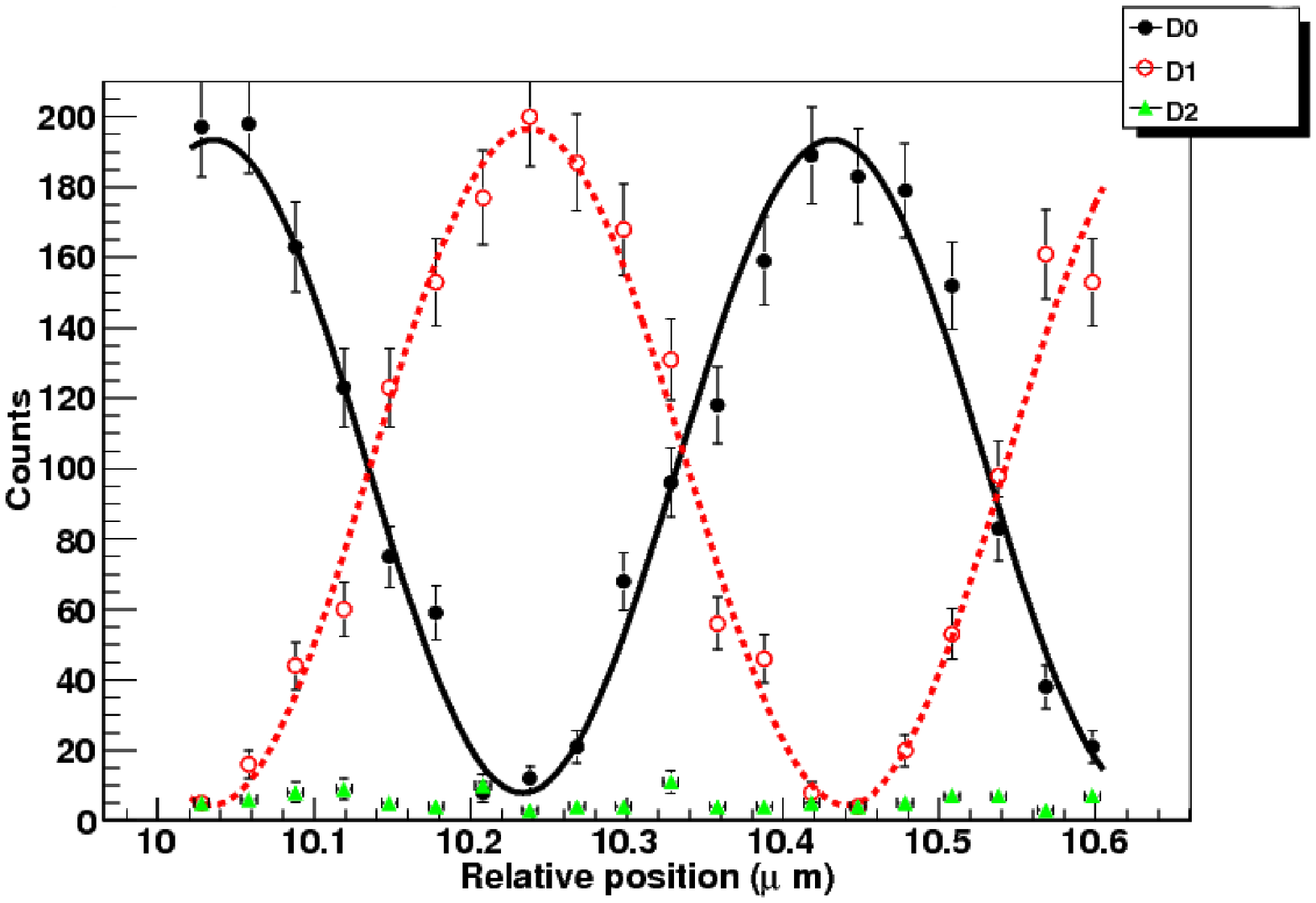}\\
   \includegraphics[scale=0.45]{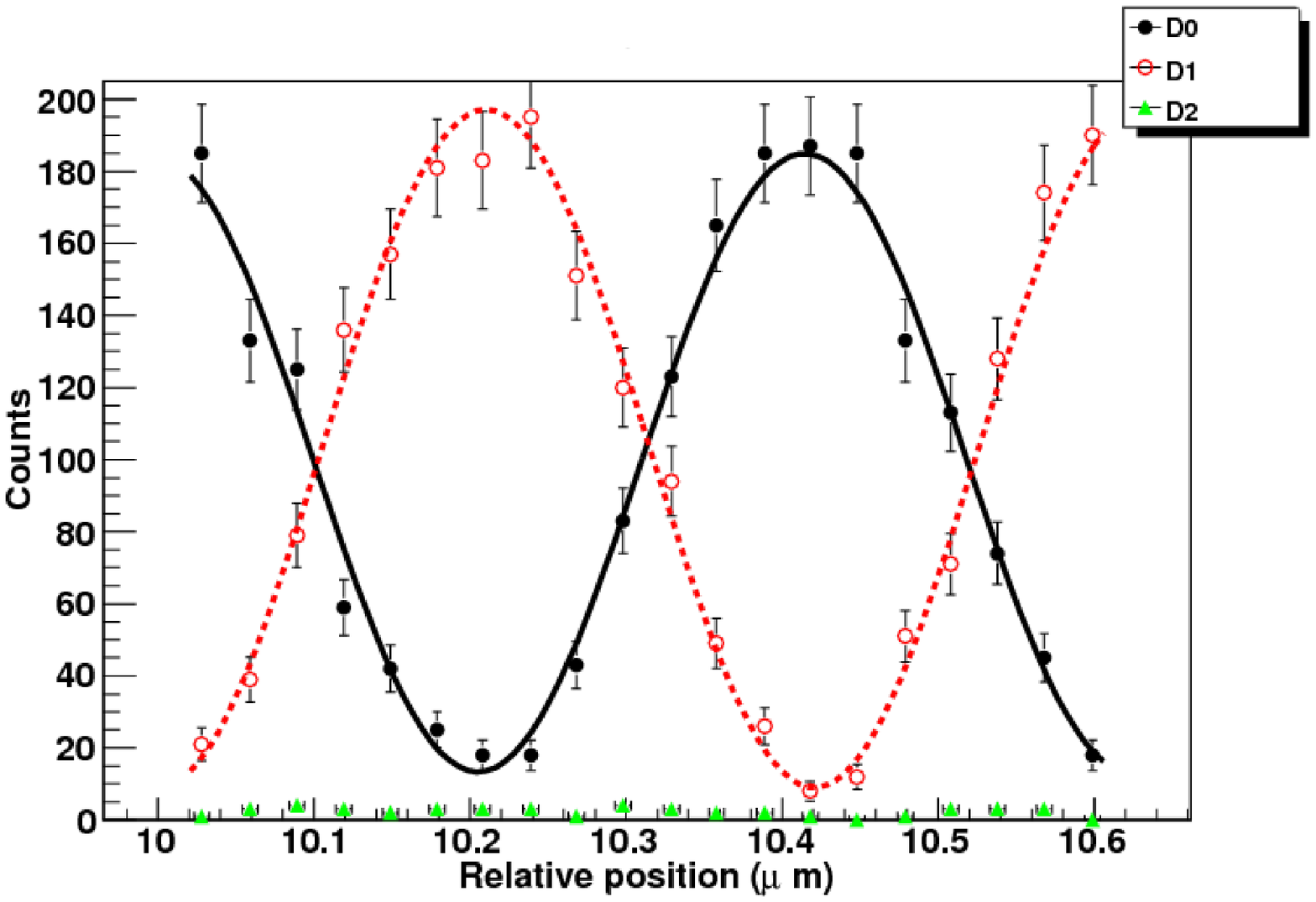}
   \end{tabular}
   \end{center}
   \caption[example2]
   { \label{fig:scan} Coincidence counts between the heralding channel and each of the MZI output detectors D0, D1 and D2 in 20 seconds acquisitions
as a function of the displacement of the prism balancing the interferometer when Alice and Bob
  use compatible sets of angles (top figure: $\{0,0\}$; bottom figure: $\{\pi/2,\pi/2\}$). For this choice of angles an interference pattern (with visibilities generally above $90\%$) can be observed in the
D0 and D1 counts and also control counts (D2) are  consistent with zero as expected.}
   \end{figure}

\begin{figure}
   \begin{center}
   \begin{tabular}{c}
   \includegraphics[scale=0.45]{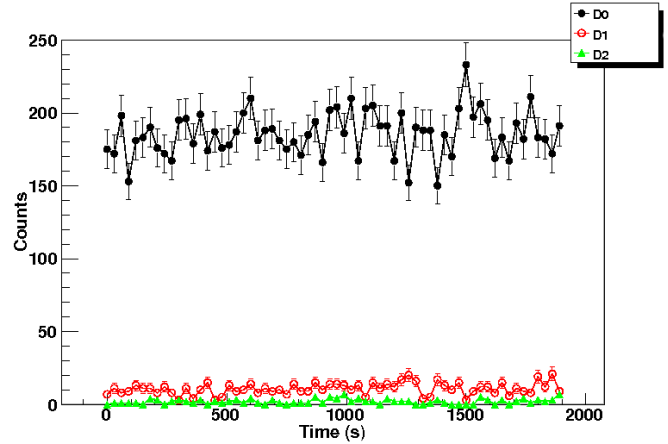}
   \end{tabular}
   \end{center}
   \caption[example2]
   { \label{fig:stab} Counting events showing the stability of the
interferometer in a half-an-hour long measurement when the balance of the two optical
paths is fixed.}
   \end{figure}


In Fig. 2 interference fringes with high visibility 
can be observed in the
coincidence counts between the heralding channel and each of the MZI output detectors D0 and D1
as a function of the displacement of the prism balancing the interferometer (within the coherence length of the signal,
which, according to the filters used, is of the order of hundreds of $\mu$m) when Alice and Bob
  use compatible sets of polarization rotation angles ($\{\theta_A,\theta_B\}=\{0,0\}$ or $\{\theta_A,\theta_B\}=\{\pi/2,\pi/2\}$). It can also be noticed that for this choice of angles the
D2 counts are consistent with zero as expected. In particular, when no rotation at all is performed ($\{0,0\}$),
the maximum visibilities are $(92 \pm 4)\%$ for D0 and $(96 \pm 4)\%$ for D1, while interference gets slightly spoiled 
for $\{\pi/2,\pi/2\}$ where the visibilities for D0 and D1 are respectively $(87 \pm 4)\%$ and $(91 \pm 4)\%$, values which, nonetheless, are sufficient for the proof of the protocol.
The uncertainty on the visibilities is obtained 
assuming a Poissonian distribution  for
the coincidence counts
.
Fig. 3 shows the stability of the
interferometer in a half-an-hour long measurement when the balance of the two optical
paths is fixed.

The performances of our key distribution process are summarized
in Table I.
Corresponding to the angles $\{0,\pi/2\}$ and $\{\pi/2,0\}$, D1 and D0 counts are approximately equal, 
as in this condition no interference should be present.
These are the events relative
to the actual transmission of information. In fact, the clicking of D1 delivers
a bit of the secret key  between the users even if no real photon travels in the
quantum channel.

In order to characterize the communication it is necessary to estimate the Quantum Bit Error Rate (QBER)
defined as the ratio between the probability for Bob to register an incorrect bit and the sum of the probabilities
of getting either a correct or an incorrect bit. In our case Bob gets an incorrect bit when D1 clicks even if Alice and Bob
use the same angle of polarization rotations and the events related to the correct transmission are those
in which D1 clicks when interference is destroyed. 
Furthermore, we notice that when Alice and Bob use complementary polarizations the amount of photons with the wrong polarization detected by D1 is effectively null when dark counts are subtracted.
We can thus define QBER as

\begin{equation}
QBER=\frac{P_{D_1,int}}{P_{D_1,int}+P_{D_1,nint}}
\end{equation}
where $P_{D_1,int}$ is the probability for D1 to register a photon when Alice's and Bob's polarization rotations are equal, such that there is (destructive) interference, and $P_{D_1,nint}$ is the analogous probability in the case in which
Alice and Bob choose different angles.

For our measurements the mean QBER is  $QBER=(12 \pm 1)\%$.
We underline that all the reported measurements are obtained without subtraction of
background and accidental counts. If we account for these contributions, the corrected
QBER value decreases noticeably to $QBER'=(7 \pm 1)\%$, 
as would be the case if more reliable detectors were used,
 such as detectors affected by a lower dark count rate
. As already mentioned, the protocol has been demonstrated absolutely secure when ideal single photon sources are employed. To address the security problems eventually raised by the practical implementation of the protocol, firstly we tested it against
possible photon-number-splitting attacks, 
i. e. we investigated the quality of of our heralded single photon source.
From the measured count rates we obtained a value of 
$g_2(0)=(7 \pm 5)*10^{-9}$, which clearly shows negligible presence of multi-photon components.
The reason for such a small value is
related to the very low level of count rates (180 maximum in 20 seconds acquisitions) at the detectors.
This is basically due to the poor coupling efficiency of the heralded source (approximately $5\%$),
 the strict spectral selection on the heralding photons (1 nm FWHM filtering with $26\%$ transmittance),
 and also because of the spatial selection at the interferometers output (we used irises as narrow as 1mm in diameter to optimize
 the visibility of the interference fringes).
Furthermore, a small temporal detection window (1 ns) was selected in correspondence  of the arrival of the heralding photon.
Because of this temporal post-selection
 we mention that unheralded photons may travel inside the channel and Eve may exploit that to get 
significant information by intercepting them.
In order to overcome this security issue, shuttered heralded single-photon sources \cite{shuttered} should be considered a valuable solution, as they present comparable performances with respect to the non-shuttered ones.
Future developments of the scheme will include shuttered sources  together with 
stabilized fiber interferometers for wider distance.

We also address the issue of robustness of the protocol against more general
attacks
by computing the difference $m=I_{AB}-I_{AE}$, where $I_{AB}$ ($I_{AE}$) is the mutual information
between Alice and Bob (Alice and Eve), in the cases of general Intercept-Resend attacks and "Time-Shift" attacks.
Following the models suggested in Ref. \cite{zhang}, one can express $m$ for the intercept-resend attack as
\be
m_{IR}=P_{D1}[1-h(\frac{P_{e1}}{P_{D1}})],
\ee
where $P_{D1}$, $P_{e1}$ are respectively the click probability and the error probability at D1 and $h(x)
$ is the binary Shannon Entropy.
Regarding the time-shift attack, where Eve exploits the non-ideality of the detectors,
one must subtract from the previous value two contributions, 
obtaining:
\be
m_{TS}=m_{IR}-\gamma-\Delta I_{AE}(\eta),
\ee
where $\gamma$ accounts for the maximum corrupted bit rate due to dark counts and $\Delta I_{AE}(\eta)=\frac{1-\eta}{2\eta}(P_{D2}-P_{e2})$  is the increment of the mutual information
between A and E due to non-unit efficiency of the detectors.

Both values 
calculated from the collected data are positive ($m_{IR}=0.23 \pm 0.04$, $m_{TS}=0.15 \pm 0.06$), 
ensuring 
the possibility of distributing a secret key \cite{gis,scar}

Altogether our results provide a satisfying proof-of-principle of the QKD scheme realized
in free-space
.
Nonetheless, recent results on the implementation of high stability fiber based Mach-Zehnder interferometers
 (over distances of the order of some km) \cite{mzfibre1, mzfibre2} 
 certify the possibility of exploiting this protocol 
in "real-life" (as well as commercial) applications.


\begin{table}[htbp]
\centering
   \begin{tabular}{|c||cccc|}
   \hline
    & $\{0,0\}$ & $\{0,\pi/2\}$ & $\{\pi/2,\pi/2\}$ & $\{\pi/2,0\}$\\
   \hline
   \hline
   $C_{D0}$ & $180 \pm 4$ & $59 \pm 2$ & $159 \pm 4$ & $59 \pm 2$\\
   \hline
   $C_{D1}$ & $7.9 \pm 0.9$ & $53 \pm 2$ & $7.2 \pm 0.9$ & $59 \pm 2$\\
   \hline
   $C_{D2}$ & $6.6 \pm 0.8$ & $85 \pm 3$ & $5.4 \pm 0.7$ & $86 \pm 3$\\
   \hline
   $V_{D0}$ & $(92 \pm 4)\%$ & $(0 \pm 4)\%$ & $(0 \pm 4)\%$ & $(87 \pm 4)\%$\\
   \hline
   $V_{D1}$ & $(96 \pm 4)\%$ & $(0 \pm 4)\%$ & $(0 \pm 4)\%$ & $(91 \pm 4)\%$\\
   \hline
   \hline
   $QBER$ & & & \hspace{-12mm}$(12 \pm 1)\%$ & \\
   \hline
   \end{tabular}
\caption{Resume of the main results in the implementation of the CQKD protocol proposed in Ref. \cite{no}.
Each column refers to a set $\{\theta_A,\theta_B\}$ of polarization rotation performed by the users and  $C_{Di}$
labels the mean coincidence counts at the $i$-th detector in acquisition of 20 seconds. $V_{D0}$, $V_{D1}$ are the visibilities of the interference fringes observed at the two outputs of the interferometer by scanning the path length difference between the two arms of the MZI. $QBER$ is the estimated quantum bit error rate
for the transmission
. }
   \end{table}

In conclusion in this paper we have presented the first experimental
demonstration of counterfactual QKD. This result, beyond its eventual
practical interest, has a huge conceptual significance since it
demonstrates for the first time as information can be transmitted
between two partners, thanks to quantum systems peculiar properties,
in a situation where no carrier has been actually transmitted
between them.

We acknowledge the support of MIQC EU project.

\end{document}